\documentclass[12pt]{article}
\usepackage{amsfonts}
\usepackage{graphicx}  
\usepackage{amssymb}
\def\be{\begin{eqnarray}}
\def\ee{\end{eqnarray}}
\begin{document}
{\LARGE
\centerline{Membranes and Consistent Quantization of Nambu Dynamics} }

\phantom{aaa} 

\centerline{\Large Cosmas Zachos }  
High Energy Physics Division,
Argonne National Laboratory, Argonne, IL 60439-4815, USA \\
\phantom{a}\qquad  \qquad \qquad  {\sl zachos@hep.anl.gov}

\begin{abstract} 
The dynamics of even topological open membranes relies on 
Nambu Brackets. Consequently, such 2p-branes can be quantized through the 
consistent quantization of the underlying Nambu dynamical structures. 
This is a summary construction relying on the methods detailed in 
refs \cite{sphere,CQNB}. 

\end{abstract}

The classical motion of topological open membranes is controlled by Nambu 
Brackets, the multilinear generalization of Poisson Brackets \cite{nambu}. 

Without loss of generality, consider first an illustrative Nambu Bracket 
(NB) dynamical law in phase space for a particle with two degrees of freedom.
Time-evolution is specified by a phase-space Jacobian,
\be
{df\over dt}= {\partial (f,L_1,L_2,L_3 ) \over \partial 
(x,p_x,y,p_y)   }\equiv \{ f, L_1,L_2,L_3 \}  
 ~.  \label{4CNB}
\ee
For an arbitrary function $f$ of phase-space variables, 
$df=\partial_l f ~dz^l$, where $z^i\equiv (x,p_x,y,p_y)$. Thus, 
this phase-space Jacobian is usually written 
symbolically as a set of Nambu 4-Brackets, 
\be
\dot{z}^l  = \{ z^l, L_1,L_2,L_3 \} . \label{motion}
\ee
The $L_i$s are arbitrary independent 
functions of the 4-d phase-space variables, and play the role of three 
``Hamiltonians",  as required in Nambu dynamics. They are manifestly 
time-invariant by the complete antisymmetry of all arguments in the 
Jacobian. 
In what action principle does this motion arise?

The action for this evolution is given by 
the analog of the Hamilton-Poincar\'{e} symplectic 
2-form ($d\omega_1=dx \wedge dp_x + dH \wedge  dt$), now 
extended to an exact 4-form \cite{estabrook,lund},
\be 
d\omega_3& =&dx \wedge dp_x \wedge dy \wedge dp_y + 
dL_1\wedge  dL_2\wedge dL_3 \wedge  dt \nonumber \\    
&=& \left (dx-\{ x , L_1,L_2,L_3 \}dt\right )\wedge \left 
 (dp_x-\{ p_x , L_1,L_2,L_3 \}dt\right ) 
\wedge \left (dy -\{ y, L_1,L_2,L_3 \}dt\right )\nonumber \\
&\phantom{+}&\wedge   \left   (dp_y-\{ p_y, L_1,L_2,L_3 \}dt \right )~.
\label{4form}
\ee 
The 4-integral of this form on an open 4-surface yields a 
3-form action 
evaluated on the 3-boundary of that surface,
\be
S=\int \Bigl ( x~ dp_x \wedge dy \wedge dp_y +  
L_1 ~dL_2\wedge dL_3 \wedge  dt\Bigr ) ~,
\ee
a Cartan integral invariant analogous to the $(2+1)$-dimensional 
$\sigma$-model  WZW topological interaction terms.  

More explicitly, in hyper-world-sheet coordinates $t, \alpha, \beta$, 
the action reads 
\be
S=\int dt d\alpha d\beta ~ \left ( {\epsilon^{ijkl}\over 4} 
 z^i \partial_t z^j 
\partial_\alpha z^k \partial_\beta z^l +  L_1 
( \partial_\alpha L_2 \partial_\beta L_3 - [\beta\alpha  ])\right ) ~ .
\label{2braneaction} 
\ee

The classical variational equations of motion resulting 
from $\delta z^i$ are
\be
0= \epsilon^{ijkl} \partial_t z^j \partial_\alpha z^k \partial_\beta z^l 
+  \partial_\alpha z^j \partial_\beta z^k ( 
\partial_i L_1 \partial_j L_2 \partial_k L_3 -[ikj] -[jik] +[jki] -[kji] 
+[kij])~.
\ee
Motion on  the membrane (along $\partial_\alpha$ or $\partial_\beta$) may be 
gauged away by virtue of $\alpha, \beta$ 
reparameterization invariance \cite{lund},
so that only the transverse motions persist in the above,
\be
\dot{z}^l =  \epsilon^{lijk} \partial_i L_1 \partial_j L_2 \partial_k L_3 ~,
\ee
which amount to  Nambu's eqns (\ref{motion}),
instead of Hamilton's equations. As for Hamilton's equations, 
manifestly, the flow in phase space is incompressible, 
since the above velocity is divergenceless,
$\partial_l \dot{z}^l=0 $ (hence Liouville's theorem).  
In form language, the ``Cauchy characteristics" are directly read off the 
4-form (\ref{4form}), whose first variation yields the above equations.

Note that, for the systems considered here, ``open membrane" is a bit of a 
misnomer, only adhered to for historical reasons. In fact, the 
actual``membrane" above is the 2-brane world-sheet 
defined on the (closed) 3-boundary of the 4-form. It is akin to a D-brane, as it 
represents the dynamics of a set of points $z^i$
which do not really influence the motion of each other:
the membrane coordinates $\alpha, \beta$ are only implicit in the  
$z^i$s; and their number may only be inferred from the above action 
whose formulation they expedited---but they do not enter explicitly in Nambu's 
equation of motion (\ref{motion}). The generalization of the above 4-form 
illustration to an arbitrary exact $p$-form, 
$d\omega_{p-1}=dz^1\wedge ... \wedge dz^p - dt\wedge 
dL_1\wedge ... \wedge dL_{p-1}$,
hence a $(p-2)$-brane, is obvious \cite{estabrook,takhtajan,matsuo,pioline}.
(Formally, it may describe $(p-2)$-branes moving in $p$-dimensional spacetimes; 
$p=2$ reduces to Hamiltonian particle mechanics in phase space 
and Poisson Brackets.)
Reference \cite{matsuo} provides a physical vortex interpretation 
for actions of this broad type. 

The quantization of such p-branes is thought to be fraught with 
complication, \cite{takhtajan,matsuo,pioline}. But, in fact, 
it {\em is} possible, since Nambu Brackets (at least in even spaces) 
may be
quantized consistently \cite{sphere,CQNB}. Let us first recall 
how ubiquitous NBs are in highly symmetric systems in phase space, and 
some of their relevant classical and quantum features.

Nambu Brackets, linear and antisymmetric in all their arguments, 
occur routinely in the classical motion specification of highly symmetric 
systems in phase-space---in fact, maximally superintegrable systems 
cannot {\em avoid} being described by NBs \cite{nutku,sphere,CQNB}. 
For $N$ degrees of freedom, hence $2N$-dimensional phase space, if there are 
extra invariants beyond the $N$ required for integrability, the 
 system is called superintegrable; at most, there are $2N-1$ algebraically 
independent integrals of motion, and then the system is called maximally 
superintegrable.

The reason is that motion is confined in phase space on the constant surfaces 
specified by these integrals, and thus their collective intersection: 
so that the phase-space velocity 
${\bf v}=(\dot{{\bf q}},\dot{{\bf p}})$ is always perpendicular to the
$2N$-d phase-space gradients $\nabla=(\partial_{\bf q}, 
\partial_{\bf p})$ of all these integrals of the motion. 
Consequently, the phase-space velocity
must be proportional to the generalized cross-product of all those gradients. 
\begin{center}  
\includegraphics[trim=0in 0.3in 0in -0.1in, width=4in]{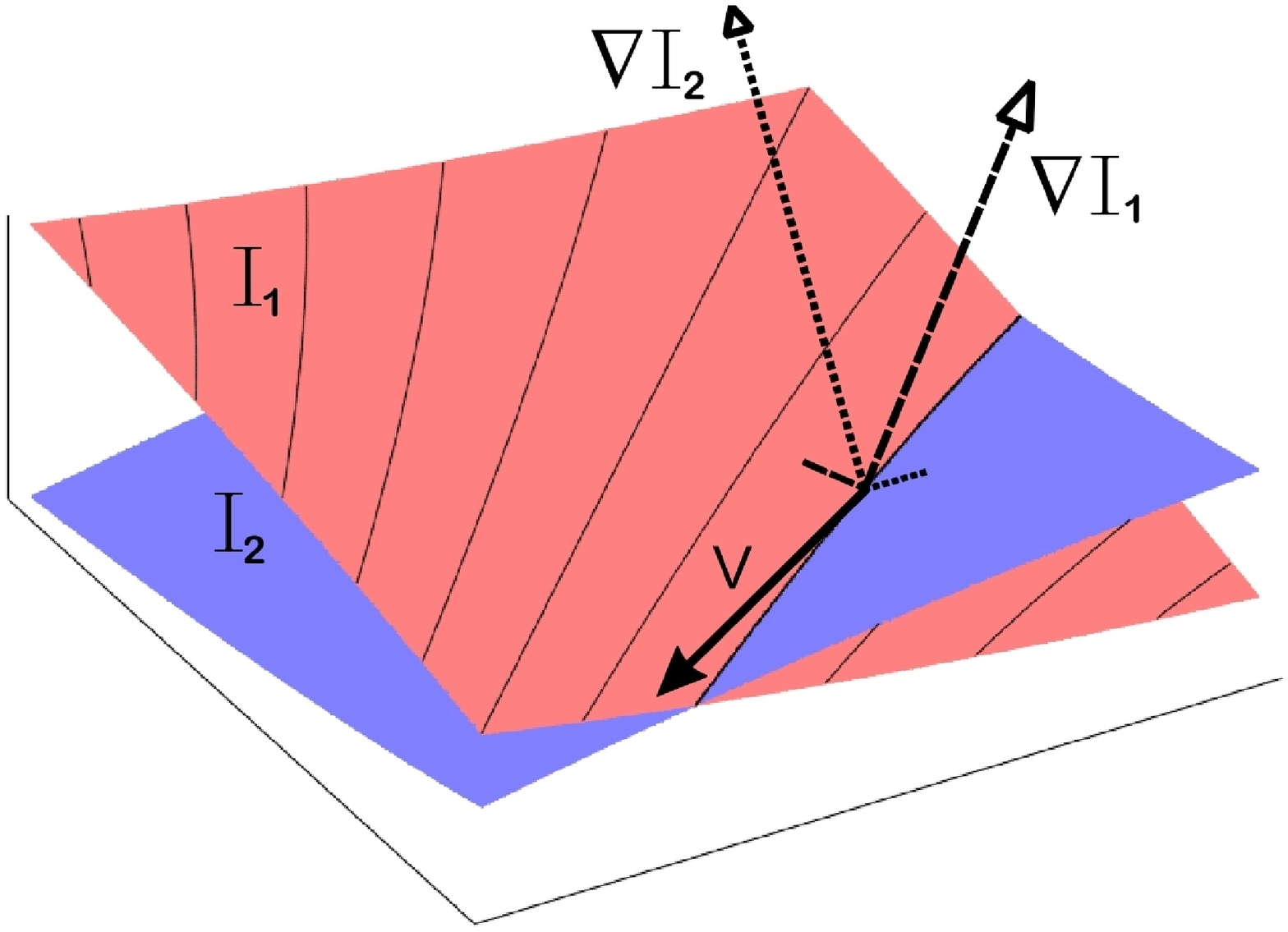}
\end{center}

It follows then that for any phase-space 
function $f({\bf q},{\bf p})$, motion is fully specified by the NB, 
\be 
{df\over dt}&=&\nabla f\cdot {\bf v} \nonumber    \\
&\propto &\partial_{i_1} f ~~ \epsilon^{i_1i_2...i_{2N}} ~ 
\partial_{i_2} L_{i_1} ... \partial_{i_{2N}}L_{2N-1} \nonumber   \\
& = & {\partial (f,L_1,...,...,L_{2N-1}) \over \partial 
(q_1,p_{1},  q_2,p_{2},  ...,q_{N},p_{N})   } \nonumber    \\
&\equiv & \{f,L_1,...,L_{2N-1}\} ~.
\ee

The proportionality constant is shown \cite{sphere,CQNB} to be time-invariant.

Thus, there is an abundance of simple classical symmetric systems controlled 
by NBs,
such as multioscillator systems, chiral models, or the free motion on 
spheres, 
even if they are also describable by Hamiltonian dynamics at the same time
\cite{sphere,CQNB,nutku,chatterjee}. 

For instance, free motion on a two-sphere $S^2$ (a beach ball) 
is specified by the quadratic Casimir Hamiltonian, 
\be
H=\frac{1}{2}\left(  L_{x}L_{x}+L_{y}L_{y}+L_{z}L_{z}\right),
\ee
where $L_{z}=xp_{y}-yp_{x}$~, $ L_{y}=-\sqrt{1-x^{2}-y^{2}}\;p_{x}$~, 
$L_{x}=\sqrt{1-x^{2}-y^{2}}\;p_{y}$, in orthogonal-projection coordinates,
where $\{L_{x},L_{y}\} =L_{z}$, ~etc. But it is also described by 
an exceptionally simple NB law,
\be
{df\over dt}= \{ f,L_x,L_y,L_z \}.   \label{S2}
\ee
Note that this is identical in form to eqn (\ref{4CNB}),
except for the additional condition that the three invariants $L_i$  now 
close under Poisson Bracketting. 

Since NBs are linear in gradients, they obey Leibniz's rule of differentiation,
\be 
\{ f(A,B) ,C,D,... \} = 
\frac{\partial f}{\partial A} \{ A ,C,D, ... \}   +
\frac{\partial f}{\partial B} \{ B ,C,D, ... \} .
\ee
Consequently, the above hamiltonian is constant, since full antisymmetry 
dictates the vanishing of the NB, 
\be 
{dH\over dt}=\left\{ { {\bf L}\cdot{\bf L} \over 2}, L_x,  L_y,  L_z\right \}=0.
\ee

In general, even rank classical 
NBs resolve into products of Poisson Brackets \cite{sphere,CQNB}. Eg, 
\be
\left\{  A,B,C,D\right\}  =\left\{  A,B\right\}
\left\{  C,D\right\}  
-\left\{
A,C\right\}  
\left\{  B,D\right\}  -\left\{  A,D\right\}  
\left\{  C,B\right\} ,  \label{Cresolution}
\ee
in comportance with full antisymmetry under 
permutations of $A,B,C,$ and $D$. 

Thus, for the above $L_i$s of the sphere, 
the 4-NB also amounts to a PB with the above Casimir Hamiltonian. 
Consequently, when the PBs of the three 2-brane invariants also close 
into $su(2)$, the 2-brane action (\ref{2braneaction}) actually yields the same 
classical equations of motion as a mere particle action,
\be  
S=\int dt \left ( \dot{x} p_x +\dot{y}  p_y - {L\cdot L \over 2} \right )~!
\ee 
 
For a less exceptional illustration (nontrivial phase-space density $V$), 
further consider 
free motion on an $N$-sphere, $S^N$, specified by a Hamiltonian, 
\be
H=\frac{1}{2}\sum_{a=1}^{N}P_{a}P_{a}+\frac{1}{4}\sum_{a,b=1}^{N}
L_{ab}L_{ab}~,
\ee
where the de Sitter momenta $P_{a}=\sqrt{1-q^2} ~ p_a$~, ~ 
for~ $a=1,\cdots,N$,  ~and $SO(N)$ rotations ~$L_{a,a+1}=q^a p_{a+1}-q^{a+1} 
p_a$~, ~~for~ $~a=1,\cdots,N-1$, 
assemble into the above quadratic Casimir Hamiltonian. However, 
equivalently to Hamilton's equations, the Nambu law of motion now yields
\be
\frac{df}{dt}= { \left(  -1\right)  ^{\left(  N-1\right)}  \over 
 P_{2}P_{3}\cdots P_{N-1} }~
\frac{\partial\left(  f,P_{1},L_{12},P_{2},L_{23},P_{3},\cdots,P_{N-1}, 
L_{N-1\;N},P_{N}\right)  }{\partial\left(  x_{1},p_{1},x_{2},p_{2},
\cdots,x_{N},p_{N}\right)}. \label{SN}
\ee

There is a standard identity often invoked for classical NBs. 
The impossibility to antisymmetrize 
more than $m$ indices in $m$-dimensional phase space, 
\be
\epsilon ^{ab....c [ i } \epsilon^{j_1 j_2 ...j_{m} ] } =0  ~,
\ee
leads to the (generalized) ``Fundamental" Identity (FI) \cite{Sahoo,takhtajan} 
\be
\{ V \{A_1,...,A_{m-1},A_m \} , A_{m+1},...,A_{2m-1}   \} +
\{A_m, V \{A_1,...,A_{m-1},A_{m+1}\},A_{m+2},...,A_{2m-1}\}  \\ 
+...+ \{A_m,...,A_{2m-2}, V \{A_1,...,A_{m-1},A_{2m-1} \} \} =
\{ A_1,...,A_{m-1},V  \{ A_m , A_{m+1},...,A_{2m-1} \}   \}, \nonumber 
\ee  
with $(2m-1)$-elements (arguments; +1, if $V$ is included), and $(m+1)$-terms. 
Note, however, that this is  {\em not}  as fundamental as the name 
suggests: it is not the generalization of the Jacobi 
Identity for PBs: it is not fully antisymmetric in all of 
its arguments, and it is not a consequence of associativity. Instead, it is 
more like a consequence of a derivation property \cite{Sahoo,CQNB}. Eg,
for the 4-NB, eqn (\ref{Cresolution}), the FI has 5 terms. Taking  
$V=1$ and further antisymmetrizing all 7 arguments of it, the FI  yields 
a $~\frac{7!}{3!4!}=35~$-term identity, the proper generalization of 
the Jacobi Identity \cite{Hanlon,Azcarraga}; it is dubbed Generalized Jacobi 
Identity (GJI) and derived and classified with its generalizations, 
extensions, and variants in ref \cite{CQNB}.

The Leibniz rule indicated for the classical NBs, amounts to an 
abstract derivation property, $\delta A=\left\{  A,B,\cdots\right\}$ 
\be 
\delta(A\mathcal{A})=A\delta\mathcal{A}+\mathcal{A}\delta A=A\left\{
\mathcal{A},B,\cdots\right\}  
+\mathcal{A}\left\{   A,B,\cdots\right\},  
\ee
and may consistently equal, eg, a time derivative of $A$, as above.
To the extent that $\delta$ and the gradients in the NB commute, as they do,
\be
\delta\{C,D,\cdots\}=\{\delta C,D,\cdots\}+\{C,\delta D,\cdots\}+\cdots\;,
\ee
and hence the FI follows:
\be 
\{\{C,D,\cdots\},B,\cdots\}=\{\{C,B,\cdots\},D,\cdots\} 
+\{C,\{D,B,\cdots\},\cdots\} +\cdots\;.
\ee
We shall see below, however, that upon quantization, this derivation 
property and the FI fail for Quantum Nambu Brackets (QNBs), 
even though the quantum generalization of the
GJI of course holds, as it encodes associativity \cite{CQNB}.

How do such NBs and hence topological membranes quantize? One normally 
seeks a consistent one-parameter ($\hbar$) deformation of the classical 
phase-space structure encountered into a quantum structure---hopefully
equivalent to the quantization of the Hamiltonian structure for the same system,
if such is available. This quantization problem, including Nambu's 
proposal for QNBs, has had an undeservedly bad reputation over the years, 
on account of top-down shortcomings. Nevertheless, none of 
the perceived consistency complications are debilitating, at least for even-NB 
systems \cite{CQNB}. In fact, such systems can be 
quantized consistently, and the results coincide with standard Hamiltonian 
quantization for specific superintegrable models which are classically 
controlled by both CNBs and, alternatively, Hamiltonian dynamics \cite{sphere}.
Reputed inconsistencies have often been addressing unsuitable (and untenable)
conditions. 

Nambu's proposal \cite{nambu} for QNBs to supplant classical NBs is 
a fully antisymmetric multilinear generalization of Heisenberg commutators 
of operators in Hilbert space corresponding conventionally to the classical 
quantities: 
\be
\left [ A, B \right ] \equiv AB-BA , 
\ee
\be
\left[ A,B,C\right] \equiv    A B C-A CB+BCA-BAC+CAB-CBA,  \label{3qnb}
\ee
\be 
\left[ A,B,C,D\right] \equiv   A [ B,C,D ] -B \left[ C,D,A\right] 
+C\left[D,A,B\right] -D \left[ A,B,C\right]   \qquad \qquad \label{4qnb}
\ee
$$
=  [A,B][C,D]+ [A,C][D,B]+ [A,D][B,C]
+[C,D][A,B]+ [D,B][A,C]+ [B,C] [A,D]~,
$$  
etc. As for classical NBs, even QNBs resolve into strings of commutators 
\cite{CQNB}, with all suitable inequivalent orderings required for full 
antisymmetry. They therefore have the proper classical 
limit: $[A_1,...,A_{2n}]\rightarrow n!(i\hbar)^n \{A_1,...,A_{2n}\}$
 as $\hbar \rightarrow 0$.
 
Full antisymmetry is the defining feature of these QNBs, but, in general,
they are not derivations. They do not satisfy the Leibniz property or the 
FI, except in special circumstances (such as the $S^2$ system below---
but not $S^N$.)  This, however, does not, in general, detract from the
consistency of the quantum equations of motion. As already indicated,
they {\em do} satisfy the proper fully antisymmetric 
Generalized Jacobi Identity \cite{Hanlon,Azcarraga,CQNB}, from associativity 
of the usual operators in conventional Hilbert space. Thus, eg, for 4-QNBs, 
\be
[[A,B,C,D],E,F,G]+ \hbox{34~signed~permutations}=0.
\ee

A difficulty with QNBs originally identified by Nambu \cite{nambu} only 
relates to odd-QNBs, but not even ones. Specifically, taking, eg,
$A\propto I$, does not force odd brackets such as (\ref{3qnb}) 
to vanish identically, which thus restricts the specification of $dA/dt$ 
by an odd QNB. In contrast, this {\em does} result in the vanishing of all even
QNBs, such as (\ref{4qnb}), by virtue of their commutator resolution. 
This consideration then is not an obstacle to a Nambu dynamical law which 
specifies $dA/dt$ to be proportional to an even QNB. 
(Odd QNBs are not proper deformations of CNBs, in general: They lack a good 
classical limit, which is at the root of the above obstacle. 
Instead, odd CNBs are reachable from larger, even QNBs like the ones discussed.
For other aspects of this odd-even dichotomy see \cite{Hanlon,Azcarraga,CQNB}.) 

For an untypical example, the quantization of (\ref{S2}) for $S^2$, is just   
\be 
\frac{df}{dt} =\frac{1}{i\hbar} [ f,H ] =
-\frac{1}{2\hbar ^{2}}\left[ f,L_{x},L_{y},L_{z}\right]\;,
\ee
as a result \cite{sphere} of the exceptional $SO(3)$-Lie-algebraic closure 
of the commutators of the quantized $L_i$'s, with (\ref{4qnb}) paralleling
the reduction of (\ref{Cresolution}). 
Thus, being equivalent to a commutator here, this QNB {\em is},
a derivation---so, here, even the Leibniz rule and the FI hold. 
Further note the good $\hbar\rightarrow 0$ limit.

But what if the ``Hamiltonians" $L_i$ do not close among themselves 
under the action of PB or commutators, respectively, eg, for the most 
general case in eqn (\ref{4CNB})? Of course, for the system to not be free, 
the $L_i$s should be algebraically independent, and the NB is the Jacobian 
of the phase-space variable change in which they act as new coordinates. 
Ignoring singular features in the effective 
phase space, one may manipulate the definitions of the $L_i$s at the classical 
level into canonical quasi-Darboux coordinates \cite{nutku,mukunda}, 
simplifying their PB structure,
but at the expense of nontrivial prefactors $V$, illustrated below.
For instance, parlaying one of the $L_i$s into an invariant combination of
the three which PB-commutes with the other two allows recasting of 
the system into a superintegrable Hamiltonian problem. Hence
the reduction of the NB into an entwined PB through (\ref{Cresolution}) 
obtains, quantized in the more typical manner 
illustrated below for $S^3$. In practical terms, however, this is more of a
strategy, rather than a specification of an automatic procedure. 
(For an apparently different approach, consider  ref \cite{nutku}.) 

For a more generic situation, consider the more typical example \cite{CQNB}
of $S^3$, where a the nontrivial prefactor $V$ may be {\em brought over to the 
left-hand side} to multiply the time derivative, classically. 
Here, the QNBs provide the correct quantization rule, 
but {\em need not} satisfy the naive Leibniz property (and FI) for 
consistency, as they are not necessarily plain derivations. Instead, 
 time derivatives are now {\em entwined inside strings of operator invariants}. 
The quantization of (\ref{SN}) for $N=3$ is then 
\be
\Biggl  [ f ,P_{1},L_{12},P_{2},L_{23},P_{3}\Biggr ] 
=3 \hbar^2 \Biggl ( P_2 [ f,H ] + [f,H ]P_2 \Biggr )+ {\cal Q}(O(\hbar^5)),
\ee 
and hence 
\be 
\Biggl  [f,P_{1},L_{12},P_{2},L_{23},P_{3}\Biggr ] =
3i\hbar^3  \frac{d}{dt} \Biggl  ( P_2 f + f P_2\Biggr ) 
+ {\cal Q}(O(\hbar^5)).
\ee
The right-hand side is not an unadorned derivation on 
$f$, so it does not impose a Leibniz rule on the left-hand side.
(Other consistency constraints are more suitable and are, of course, 
satisfied, including the GJI.) 
${\cal Q}(O(\hbar^5))$ is a subdominant nested commutator 
``quantum rotation" \cite{CQNB}.
Solving for $df/dt$ may be more challenging technically, 
but the formulation is still equivalent to the standard Hamiltonian 
quantization of this problem \cite{CQNB}.

As a hypothetical wisecrack, ignoring ${\cal Q}$ for the sake of argument, 
one might imagine solving the above difficult Jordan-Kurosh spectral 
problem, 
\be
\Biggl [ f,P_{1},L_{12},P_{2},L_{23},P_{3} \Biggr ] \sim 
3i\hbar^3  \Biggl  ( P_2 \frac{df}{dt} + \frac{df}{dt} P_2\Biggr ),
\ee
eg, assuming invertibility of $P_2$. Formally,  the resolvent of the above 
would yield 
\be
3i\hbar^3 \frac{df}{dt} \sim \sum_{n=0}^{\infty}
(-P_2)^n  \Biggl [ f,P_{1},L_{12},P_{2},L_{23},P_{3}\Biggr ] (P_2)^{-n-1}, 
\ee
so that the right-hand side would then furnish a vision of a different 
``quantum bracket", with the proper full antisymmetry,  which now {\em is} a 
derivation, but at evident sacrifice to simplicity and generality. 

This general method of quantization, successful in a large number of 
systems, is detailed in \cite{CQNB}. Essentially, as suggested by the 
classical NB, the commutator resolution of a suitably 
chosen QNB parallels the classical combinatorics to yield a commutator 
of the operator unknown with the hamiltonian (hence its time derivative), 
entwined with invariants. 

In summary,  Quantum Nambu Brackets are consistent and describe the 
quantum behavior of superintegrable systems equivalently to standard 
Hamiltonian quantization, and thus serve to guide quantization of more general 
even-dimensional topological membranes. 

\phantom{..}

\noindent{\bf Acknowledgement} ~~~Obligation is recorded to T Curtright for 
initiating appreciation of ref \cite{estabrook} and trenchant comments,
 and A Polychronakos for useful questions. 
This work was supported by the US Department of Energy, 
Division of High Energy Physics, Contract W-31-109-ENG-38.

\end{document}